\documentclass[11pt,letter]{article}

\usepackage{amsmath,amscd,fancyhdr,graphics,pifont}
\usepackage{floatflt,subfigure,epsfig,moreverb,multicol,lscape}
\usepackage{supertabular,tabularx,multirow,bar,amssymb}
\usepackage{theorem,psfrag,afterpage,curves,epic, bbm}
\usepackage[english]{babel}

\newcommand{\ignore}[1]{}

\newcommand{\dmins}{d_{\mathrm{min}}}

\newcommand{\matr}[1]{\mathbf{#1}}
\newcommand{\vect}[1]{\mathbf{#1}}

\newcommand{\set}[1]{\mathcal{#1}}

\newcommand{\GF}[1]{\mathbb{F}_{#1}}

\newcommand{\defeq}{\triangleq}

\newcommand{\Z}{\mathbb{Z}}

\newcommand{\vx}{\vect{x}}

\newcommand{\vy}{\vect{y}}

\newcommand{\mDelta}{\boldsymbol{\Delta}}

\newcommand{\tr}{\mathsf{T}}

\newcommand{\HD}[2]{\mathcal{D}_{#1}^{(#2)}}
\newcommand{\iL}{\mathsf{L}}
\newcommand{\iX}{\mathsf{X}}

\makeatletter
  \def\revddots{\mathinner{\mkern1mu\raise\p@
  \vbox{\kern7\p@\hbox{.}}\mkern2mu
  \raise4\p@\hbox{.}\mkern2mu\raise7\p@\hbox{.}\mkern1mu}}
  \makeatother

\newtheorem{Lemma}{Lemma}

\newtheorem{Proposition}[Lemma]{Proposition}
\newtheorem{Corollary}[Lemma]{Corollary}

\theoremstyle{plain}

\newtheorem{PreDefinition}[Lemma]{{\textbf{Definition}}}
  \newenvironment{Definition}%
    {\begin{PreDefinition}}{\hfill$\square$\end{PreDefinition}}

\theoremstyle{plain}

\newtheorem{PreRemark}[Lemma]{{\textbf{Remark}}}
    {\begin{PreRemark}\upshape}{\hfill$\square$\end{PreRemark}}

\newtheorem{PreExample}[Lemma]{{\textbf{Example}}}
  \newenvironment{Example}%
    {\begin{PreExample}\upshape}{\hfill$\square$\end{PreExample}}

\newenvironment{Proof}%
  {\noindent \emph{Proof:}}{\hfill$\square$}

\begin{document}

\title{An Explicit Construction of Universally Decodable Matrices%
  \footnote{The first author was supported by NSF Grants ATM-0296033 and DOE
            SciDAC and by ONR Grant N00014-00-1-0966. The second author was
            supported by NSF Grant TF-0514801. Some preliminary results
            discussed here also appear in \cite{Ganesan:Boston:05:1}.}}

\author{Pascal O.~Vontobel and Ashwin Ganesan%
  \thanks{ECE Department, University of Wisconsin-Madison,
          1415 Engineering Drive Madison, WI 53706, USA. 
          Email: \texttt{vontobel@ece.wisc.edu}, \texttt{ganesan@cae.wisc.edu}.
          }}

\date{}

\maketitle

\begin{abstract}
  Universally decodable matrices can be used for coding purposes when
  transmitting over slow fading channels. These matrices are parameterized by
  positive integers $L$ and $n$ and a prime power $q$. Based on Pascal's
  triangle we give an explicit construction of universally decodable matrices
  for any non-zero integers $L$ and $n$ and any prime power $q$ where $L \leq
  q+1$. This is the largest set of possible parameter values since for any
  list of universally decodable matrices the value $L$ is upper bounded by
  $q+1$, except for the trivial case $n = 1$. For the proof of our
  construction we use properties of Hasse derivatives, and it turns out that
  our construction has connections to Reed-Solomon codes, Reed-Muller codes,
  and so-called repeated-root cyclic codes. Additionally, we show how
  universally decodable matrices can be modified so that they remain
  universally decodable matrices.
\end{abstract}


\section{Introduction}
\label{sec:introduction:1}

\begin{figure}[h]
  \begin{center}
    \epsfig{file=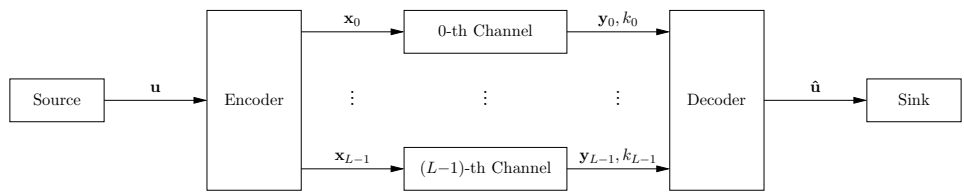, width=12cm}
  \end{center}
  \caption{Communication system with $L$ parallel channels.}
  \label{fig:communication:system:1}
\end{figure}

Let $L$ and $n$ be non-zero integers, let $q$ be a prime power, let $[M] 
\defeq \{ 0, \ldots, M-1 \}$ for any positive integer $M$, and let $[M] \defeq 
\{ \ \}$ for any non-positive integer $M$. While studying slow fading 
channels (c.f.~e.g.~\cite{Tse:Viswanath:05:1}), Tavildar and
Viswanath~\cite{Tavildar:Viswanath:05:1} introduced the communication system
shown in Fig.~\ref{fig:communication:system:1} which works as follows. An
information (column) vector $\vect{u} \in \GF{q}^n$ is encoded into vectors
$\vx_{\ell} \defeq \matr{A}_{\ell} \cdot \vect{u} \in
\GF{q}^n$, $\ell \in [L]$, where $\matr{A}_0, \ldots, \matr{A}_{L-1}$ are $L$
matrices over $\GF{q}$ of size $n \times n$. Upon sending $\vx_{\ell}$ over
the $\ell$-th channel we receive $\vy_{\ell} \in (\GF{q} \cup \{ ? \})^n$,
where the question mark denotes erasures. The channels are such that the
received vectors $\vect{y}_1, \ldots, \vect{y}_{L-1}$ can be characterized as
follows: there are integers $k_0, \ldots, k_{L-1}$, $0 \leq k_{\ell} \leq n$,
$\ell \in [L]$ (that can vary from transmission to transmission) such that the
first $k_{\ell}$ entries of $\vect{y}_{\ell}$ are non-erased and agree with
the corresponding entries of $\vect{x}_{\ell}$ and such that the last $n -
k_{\ell}$ entries of $\vect{y}_{\ell}$ are erased.

Based on these non-erased entries we would like to reconstruct $\vect{u}$. The
obvious decoding approach works as follows: construct a $(\sum_{\ell \in [L]}
k_{\ell}) \times n$-matrix $\matr{A}$ that stacks the $k_0$ first rows of
$\matr{A}_0$, $\ldots$, the $k_{L-1}$ first rows of $\matr{A}_{L-1}$; then
construct a length-$(\sum_{\ell \in [L]} k_{\ell})$ vector $\vect{y}$ that
concatenates the $k_0$ first entries of $\matr{y}_0$, $\ldots$, the
$k_{L-1}$ first entries of $\matr{A}_{L-1}$; finally, the vector $\vect{\hat
u}$ is given as the solution of the linear equation system $\matr{A}
\cdot \vect{\hat u} = \vect{y}$. Since $\vect{u}$ is arbitrary in $\GF{q}^n$,
a necessary condition for successful decoding is that $\sum_{\ell \in [L]}
k_{\ell} \geq n$. Because we would like to be able to decode correctly for all
$L$-tuples $(k_0, \ldots, k_{L-1})$ that satisfy this necessary condition, we
must guarantee that the matrix $\matr{A}$ has full rank for all possible
$L$-tuples $(k_0, \ldots, k_{L-1})$ with $\sum_{\ell \in [L]} k_{\ell} \geq
n$. Matrices that fulfill this condition are called universally decodable
matrices (UDMs).

Given this setup there are two immediate questions. First, for what values of
$L$, $n$, and $q$ do such matrices exist?  Secondly, how can one construct
such matrices? In~\cite{Tavildar:Viswanath:05:1} a construction is given for
$L = 3$, any $n$, and $q = 2$. Doshi~\cite{Doshi:05:1} gave a construction for
$L = 4$, $n = 3$, and $q = 3$ and conjectured a construction for $L = 4$, $n$
any power of $3$, and $q = 3$. Ganesan and Boston~\cite{Ganesan:Boston:05:1}
showed that for any $n \geq 2$ the value $L$ is upper bounded by $L \leq
q+1$. In this paper we will give an explicit construction that works for any
positive integers $L$ and $n$ and any prime power $q$ as long as $L \leq q+1$,
in other words, this construction achieves for any $n \geq 2$ and any prime
power $q$ the above-mentioned upper bound on $L$. To the best of our knowledge
this is the first construction of universally decodable matrices that covers
all possible parameter values. As a side result, our construction shows that
the above-mentioned conjecture is indeed true.

The above problem is reminiscent of the following well-know problem. An
information vector $\vect{u} \in \GF{q}^{n}$ is encoded into the vector
$\vect{x} \defeq \matr{G} \cdot \vect{u} \in \GF{q}^{n'}$ where $\matr{G}$ is
an $n' \times n$-matrix $\matr{G}$. Upon sending $\vect{x}$ over an erasure
channel we receive $\vect{y} \in (\GF{q} \cap \{ ? \})^{n'}$: the $i$-th entry
of $\vect{y}$ is either equal to the $i$-th entry of $\vect{x}$ or equal to
the question mark. Since $\vect{u} \in \GF{q}^n$ is arbitrary, a necessary
condition for successful reconstruction is that the number of non-erased
entries is at least $n$. Because we would like to be able to decode
successfully whenever the number of non-erased entries is at least $n$ this
implies that all sub-matrices of $\matr{G}$ of size $n \times n$ must have
full rank. This problem is well-studied and leads to so-called
maximum-distance separable (MDS) codes like Reed-Solomon
codes~\cite{MacWilliams:Sloane:77, Blahut:02:1}. As was noted
in~\cite[Sec.~4.5.5]{Tavildar:Viswanath:05:1}, for $L$, $n$, and $q$ such that
$q \geq Ln-1$ the problem of constructing UDMs can be reduced to the problem
of constructing MDS codes. However, the required field size ($q \geq Ln-1$) is
much larger than the field size that is required by our UDMs construction ($q
\geq L-1$).

The paper is structured as follows. In Sec.~\ref{sec:udm:1} we properly define
UDMs and in Sec.~\ref{sec:modifying:udms:1} we show how UDMs can be modified
to obtain new UDMs. Sec.~\ref{sec:explicit:construction:udms:1} is the main
section where an explicit construction of UDMs is presented. In
Sec.~\ref{sec:conclusions:1} we offer some conclusions,
Sec.~\ref{sec:proofs:1} contains the longer proofs, and
Sec.~\ref{sec:hasse:derivative:1} collects some results on Hasse derivatives
which are the main tool for the proof of our UDMs construction.

\section{Universally Decodable Matrices}
\label{sec:udm:1}

The notion of universally decodable matrices (UDMs) was introduced by Tavildar
and Viswanath~\cite{Tavildar:Viswanath:05:1}. Before we give the definition of
UDMs, let us agree on some notation. For any positive integer $n$, we let
$\matr{I}_n$ be the $n \times n$ identity matrix and we let $\matr{J}_n$ be
the $n \times n$ matrix where all entries are zero except for the
anti-diagonal entries that are equal to one. Row and column indices of
matrices will always be counted from zero on and the entry in the $i$-th row
and $j$-th column of $\matr{A}$ will be denoted by
$[\matr{A}]_{i,j}$. Similarly, indices of vectors will be counted from zero on
and the $i$-th entry of $\vect{a}$ will be denoted by $[\vect{a}]_i$. For any
positive integer $L$ and any non-negative integer $n$ we define the sets
\begin{align*}
  \set{K}^{= n}_L
    &\defeq
       \left\{
         (k_0, \ldots, k_{L-1})
         \ | \
         0 \leq k_{\ell} \leq n, 
         \ell \in [L], \ 
         \sum_{\ell \in [L]} k_{\ell} = n
       \right\}, \\
  \set{K}^{\geq n}_L
    &\defeq
       \left\{
         (k_0, \ldots, k_{L-1})
         \ | \ 0
         \leq k_{\ell} \leq n,
         \ell \in [L], \ 
         \sum_{\ell \in [L]} k_{\ell} \geq n
       \right\}.
\end{align*}

\begin{Definition}
  \label{def:udms:1}

  Let $n$ and $L$ be some positive integers and let $q$ be a prime power. The
  $L$ matrices $\matr{A}_0, \ldots, \matr{A}_{L-1}$ over $\GF{q}$ and size $n
  \times n$ are $(L,n,q)$-UDMs, or simply UDMs, if for every $(k_0, \ldots,
  k_{L-1}) \in \set{K}^{\geq n}_L$ they fulfill the UDMs condition which
  says that the $( \sum_{\ell \in [L]} k_{\ell} ) \times n$ matrix composed of
  the first $k_0$ rows of $\matr{A}_0$, the first $k_1$ rows of $\matr{A}_1$,
  $\ldots$, the first $k_{L-1}$ rows of $\matr{A}_{L-1}$ has full rank.
\end{Definition}

We list some immediate consequences of the above definition.
\begin{itemize}

  \item To assess that some matrices $\matr{A}_0, \ldots, \matr{A}_{L-1}$ are
    UDMs, it is sufficient to check the UDMs condition only for every $(k_0,
    \ldots, k_{L-1}) \in \set{K}^{=}_{L,n}$. There are ${ n+L-1 \choose L-1 }$
    such $L$-tuples.

  \item If the matrices $\matr{A}_0, \ldots, \matr{A}_{L-1}$ are UDMs then
    these matrices are all invertible.

  \item If the matrices $\matr{A}_0, \ldots, \matr{A}_{L-1}$ are
    $(L,n,q)$-UDMs then they are $(L,n,q')$-UDMs for any $q'$ that is a power
    of $q$.

  \item Let $\sigma$ be any permutation of $[L]$. If the matrices $\matr{A}_0,
    \ldots, \matr{A}_{L-1}$ are $(L,n,q)$-UDMs then the matrices
    $\matr{A}_{\sigma(0)}, \ldots, \matr{A}_{\sigma(L-1)}$ are also
    $(L,n,q)$-UDMs.

  \item If the matrices $\matr{A}_0, \ldots, \matr{A}_{L-1}$ are
    $(L,n,q)$-UDMs then the matrices $\matr{A}_0, \ldots, \matr{A}_{L'-1}$ are
    $(L',n,q)$-UDMs for any positive $L'$ with $L' \leq L$.

  \item If the matrices $\matr{A}_0, \ldots, \matr{A}_{L-1}$ are
    $(L,n,q)$-UDMs and $\matr{B}$ is an invertible $n \times n$-matrix over
    $\GF{q}$ then the matrices $\matr{A}_0 \cdot \matr{B}, \ldots,
    \matr{A}_{L-1} \cdot \matr{B}$ are $(L,n,q)$-UDMs. Without loss of
    generality, we can therefore assume that $\matr{A}_0 = \matr{I}_n$.

  \item For $n = 1$ we see that for any positive integer $L$ and any prime
    power $q$, the $L$ matrices $(1), \ldots, (1)$ are
    $(L,n{=}1,q)$-UDMs. Because of the trivial-ness of the case $n = 1$, the
    rest of the paper focuses on the case $n \geq 2$.

\end{itemize}

\begin{Example}
  \label{ex:udm:1}

  Let $n$ be any positive integer, let $q$ be any prime power, and let $L
  \defeq 2$. Let $\matr{A}_0 \defeq \matr{I}_n$ and let $\matr{A}_1 \defeq
  \matr{J}_n$. It can easily be checked that $\matr{A}_0, \matr{A}_1$ are
  $(L{=}2,n,q)$-UDMs. Indeed, let for example $n \defeq 5$. We must check that
  for any non-negative integers $k_1$ and $k_2$ such that $k_1 + k_2 = 5$ the
  UDMs condition is fulfilled. E.g.~for $(k_1, k_2) = (3, 2)$ we must show
  that the matrix
  \begin{align*}
    \begin{pmatrix}
      1 & 0 & 0 & 0 & 0 \\
      0 & 1 & 0 & 0 & 0 \\
      0 & 0 & 1 & 0 & 0 \\
      0 & 0 & 0 & 0 & 1 \\
      0 & 0 & 0 & 1 & 0 
    \end{pmatrix}
  \end{align*} 
  has rank $5$, which can easily be verified.
\end{Example}

\begin{Example}
  \label{ex:udm:2}

  In order to give the reader a feeling how UDMs might look like for $L > 2$,
  we give here a simple example for $L = 4$, $n = 3$, and $q = 3$, namely
  \begin{align*}
    \matr{A}_0
      &= \begin{pmatrix}
           1 & 0 & 0 \\
           0 & 1 & 0 \\
           0 & 0 & 1
         \end{pmatrix},
         \quad
    \matr{A}_1
      &= 
         \begin{pmatrix}
           0 & 0 & 1 \\
           0 & 1 & 0 \\
           1 & 0 & 0
         \end{pmatrix},
         \quad
    \matr{A}_2
      &= 
         \begin{pmatrix}
           1 & 1 & 1 \\
           0 & 1 & 2 \\
           0 & 0 & 1 
         \end{pmatrix},
         \quad
    \matr{A}_3
      &= 
         \begin{pmatrix}
           1 & 2 & 1 \\
           0 & 1 & 1 \\
           0 & 0 & 1 
         \end{pmatrix}.
  \end{align*}
  One can verify that for all $(k_0, k_1, k_2, k_3) \in \set{K}^{= 3}_4$
  (there are $20$ such four-tuples) the UDMs condition is fulfilled and hence
  the above matrices are indeed UDMs. For example, for $(k_0,k_1,k_2,k_3) =
  (0,0,3,0)$, $(k_0,k_1,k_2,k_3) = (0,0,1,2)$, and $(k_0,k_1,k_2,k_3) =
  (1,1,0,1)$ the UDMs condition means that we have to check if the matrices
  \begin{align*}
    \begin{pmatrix}
      1 & 1 & 1 \\
      0 & 1 & 2 \\
      0 & 0 & 1 
    \end{pmatrix},
    \quad
    \begin{pmatrix}
      1 & 1 & 1 \\
      1 & 2 & 1 \\
      0 & 1 & 1
    \end{pmatrix},
    \quad
    \begin{pmatrix}
      1 & 0 & 0 \\
      0 & 0 & 1 \\
      1 & 2 & 1
    \end{pmatrix}
  \end{align*}
  have rank $3$, respectively, which is indeed the case. Before concluding
  this example, let us remark that the above UDMs are the same UDMs that
  appeared in~\cite{Doshi:05:1}
  and~\cite[Sec.~4.5.4]{Tavildar:Viswanath:05:1}.
\end{Example}

\section{Modifying UDMs}
\label{sec:modifying:udms:1}

\begin{Lemma}
  \label{lemma:replace:and:add:lines:1}
  
  Let $\matr{A}_0, \ldots, \matr{A}_{L-1}$ be $(L,n,q)$-UDMs. For any $\ell
  \in [L]$ and $i \in [n]$ we can replace the $i$-th row of $\matr{A}_{\ell}$
  by any non-zero multiple of itself without violating any UDMs
  condition. Moreover, for any $\ell \in [L]$ and $i, i' \in [n]$, $i > i'$,
  we can add any multiples of the $i'$-th row of $\matr{A}_{\ell}$ to the
  $i$-th row of $\matr{A}_{\ell}$ without violating any UDMs condition. More
  generally, the matrix $\matr{A}_{\ell}$ can be replaced by $\matr{C}_{\ell}
  \cdot \matr{A}_{\ell}$ without violating any UDMs condition, where
  $\matr{C}_{\ell}$ is an arbitrary lower triangular $n \times n$-matrix over
  $\GF{q}$ with non-zero diagonal entries.
\end{Lemma}

\begin{Proof}
  Follows from well-known properties of determinants.
\end{Proof}

\begin{Lemma}
  \label{lemma:tensoring:1}
  
  Let $\matr{A}_0, \ldots, \matr{A}_{L-1}$ be $(L,n,q)$-UDMs for which we know
  that the tensor powers $\matr{A}_0^{\otimes m}, \ldots,
  \matr{A}_{L-1}^{\otimes m}$ are $(L,n^m,q)$-UDMs for some positive integer
  $m$. For all $\ell \in [L]$, let $\matr{A}'_{\ell} \defeq \matr{A}_{\ell}
  \cdot \matr{B}$, where $\matr{B}$ is an arbitrary invertible $n \times n$
  matrix over $\GF{q}$. Then $\matr{A}'_0, \ldots, \matr{A}'_{L-1}$ are
  $(L,n,q)$-UDMs and $(\matr{A}'_0)^{\otimes m}, \ldots,
  (\matr{A}'_{L-1})^{\otimes m}$ are $(L,n^m,q)$-UDMs. On the other hand, if
  for all $\ell \in [L]$ we define $\matr{A}'_{\ell} \defeq \matr{C}_{\ell}
  \cdot \matr{A}_{\ell}$, where $\matr{C}_{\ell}$, $\ell \in [L]$, are
  lower-triangular matrices with non-zero diagonal entries, then $\matr{A}'_0,
  \ldots, \matr{A}'_{L-1}$ are $(L,n,q)$-UDMs and $(\matr{A}'_0)^{\otimes m},
  \ldots, (\matr{A}'_{L-1})^{\otimes m}$ are $(L,n^m,q)$-UDMs.
\end{Lemma}

\begin{Proof}
  This follows from the sixth comment after Def.~\ref{def:udms:1}, from
  Lemma~\ref{lemma:replace:and:add:lines:1}, and by using a well-known
  property of tensor products, namely that $(\matr{M}_{1} \cdot
  \matr{M}_2)^{\otimes m} = \matr{M}_1^{\otimes m} \cdot \matr{M}_2^{\otimes
  m}$ for any compatible matrices $\matr{M}_1$ and $\matr{M}_2$. Note that 
  $\matr{C}_{\ell}^{\otimes m}$ is a lower-triangular matrix with non-zero
  diagonal entries for all $\ell \in [L]$ and positive integers $m$.
\end{Proof}

\begin{Lemma}
  \label{lemma:reversed:order:1}
  
  Let $\matr{A}_0, \ldots, \matr{A}_{L-1}$ be $(L,n,q)$-UDMs. Then there exist
  matrices $\matr{A}'_0, \ldots, \matr{A}'_{L-1}$ that are $(L,n,q)$-UDMs and
  where for all $\ell' \in \lfloor L/2 \rfloor$ the matrix
  $\matr{A}'_{2\ell'+1}$ is the same as $\matr{A}'_{2\ell'}$ except that the
  rows are in reversed order, i.e.~$\matr{A}_{2\ell'+1} = \matr{J}_n \cdot
  \matr{A}'_{2\ell'}$.
\end{Lemma}

\begin{Proof}
  See Sec.~\ref{sec:proof:lemma:reversed:order:1}.
\end{Proof}

\mbox{}

From Lemma~\ref{lemma:reversed:order:1} we see that when considering
$(L,n,q)$-UDMs $\matr{A}_0, \ldots, \matr{A}_{L-1}$ we can without loss of
generality assume that $\matr{A}_0 = \matr{I}_n$ and that $\matr{A}_1 =
\matr{J}_n$. Indeed, if $\matr{A}_0$ and $\matr{A}_1$ are not of this form
then the algorithm in the proof of Lemma~\ref{lemma:reversed:order:1} allows
us to replace these two matrices by two matrices where $\matr{A}_1$ is the
same as $\matr{A}_0$ except that the rows are in reversed order,
i.e.~$\matr{A}_1 = \matr{J}_n \cdot \matr{A}_0$. Let $\matr{B}
\defeq \matr{A}_0^{-1}$. Replacing for all $\ell \in [L]$ the matrix
$\matr{A}_{\ell}$ by the matrix $\matr{A}_{\ell} \cdot \matr{B}$ we obtain the
desired result.

\begin{Lemma}
  \label{lemma:reduction:1}

  Let the matrices $\matr{A}_0, \ldots, \matr{A}_{L-1}$ be $(L,n,q)$-UDMs with
  $\matr{A}_0 = \matr{I}_n$ and $\matr{A}_1 = \matr{J}_n$. The matrices
  $\matr{A}'_0, \ldots, \matr{A}'_{L-1}$ are $(L,n{-}1,q)$-UDMs if
  $\matr{A}'_{\ell}$ is obtained as follows from $\matr{A}_{\ell}$: if $\ell =
  1$ then delete the first column and last row of $\matr{A}_{\ell}$, otherwise
  delete the last column and last row of $\matr{A}_{\ell}$.
\end{Lemma}

\begin{Proof}
  See Sec.~\ref{sec:proof:lemma:reduction:1}.
\end{Proof}

\begin{Lemma}[\cite{Ganesan:Boston:05:1}]
  \label{lemma:maximal:L:1}
  
  The above results imply the following bound: if $n \geq 2$ then
  $(L,n,q)$-UDMs can only exist for $L \leq q+1$. (Note that this upper bound
  on $L$ is independent of $n$ as long as $n \geq 2$.)
\end{Lemma}

\begin{Proof}
  See Sec.~\ref{sec:proof:lemma:maximal:L:1}.
\end{Proof}

\section{An Explicit Construction of UDMs}
\label{sec:explicit:construction:udms:1}

We introduce some conventions and notations that will be used in this
section. First, whenever necessary we use the natural mapping of the integers
into the prime subfield\footnote{When $q = p^s$ for some prime $p$ and some
positive integer $s$ then $\GF{p}$ is a subfield of $\GF{q}$ and is called the
prime subfield of $\GF{q}$. $\GF{p}$ can be identified with the integers where
addition and multiplication are modulo $p$.} of $\GF{q}$. Secondly, we define
the binomial coefficients in the usual way, i.e.~for any integers $a$ and $b$
we let
\begin{align}
  {a \choose b}
    &\defeq
       \frac{a \cdot (a-1) \cdots (a - b + 2) \cdot (a - b + 1)}
            {b \cdot (b-1) \cdots 2 \cdot 1}
       \label{eq:binomial:coefficient:1}
\end{align}
if $b$ is positive, ${a \choose b} \defeq 1$ if $b$ equals zero, and ${a
\choose b} \defeq 0$ if $b$ is negative. One can check that for any 
integers $a$ and $b$ this yields ${a \choose b} \in \Z$ and the well-known
relationship ${a \choose b} = {a-1 \choose b-1} + {a-1 \choose b}$ among
different binomial coefficients.\footnote{It is probably the best to think of
${a \choose b}$ as a function $\Z \times \Z \to \Z$. The relationship ${a
\choose b} = {a-1 \choose b-1} + {a-1 \choose b}$ holds obviously over any
$\GF{q}$ where $q$ is a prime power. Note that this is the only fact we need
about binomial coefficients, i.e.~we do not need the ``internal structure'' on
the right-hand side of of~\eqref{eq:binomial:coefficient:1}.}

\begin{Proposition}
  \label{prop:udms:construction:1}

  Let $n$ be some positive integer, let $q$ be some prime power, and let
  $\alpha$ be a primitive element in $\GF{q}$, i.e.~$\alpha$ is an $(q-1)$-th
  primitive root of unity. If $L \leq q+1$ then the following $L$ matrices
  over $\GF{q}$ of size $n \times n$ are $(L, n, q)$-UDMs:
  \begin{align*}
    &
    \matr{A}_0
       \defeq \matr{I}_n,
    \quad
    \matr{A}_1
       \defeq \matr{J}_n,
    \quad
    \matr{A}_{2},
    \quad
    \ldots,
    \quad
    \matr{A}_{L-1}, \\
    &
    \text{ where } \
    [\matr{A}_{\ell+2}]_{i,t}
       \defeq {t \choose i} \alpha^{\ell (t-i)}, \ 
              (\ell,i,t) \in [L-2] \times [n] \times [n].
  \end{align*}
  Note that ${t \choose i}$ is to be understood as follows: compute ${t
  \choose i}$ over the integers and apply only then the natural mapping to
  $\GF{q}$.
\end{Proposition}

\begin{Proof}
  See Sec.~\ref{sec:proof:prop:udms:construction:1}. However, before looking
  at the proof we recommend to first study Ex.~\ref{ex:udms:construction:1}
  and secondly to familiarize oneself with Hasse derivatives,
  cf.~Sec.~\ref{sec:hasse:derivative:1}. Especially
  Lemma~\ref{lemma:evaluation:1} and Cor.~\ref{cor:evaluation:2} in
  Sec.~\ref{sec:hasse:derivative:1} are interesting since they will provide the
  key for proving the proposition.
\end{Proof}

\begin{Example}
  \label{ex:udms:construction:1}

  For $n \defeq 3$, $p \defeq 3$, and $\alpha \defeq 2$, we obtain the $L = 3
  + 1 = 4$ matrices that were shown in Ex.~\ref{ex:udm:2}. Note that
  $\matr{A}_3$ is nearly the same as $\matr{A}_2$: it differs only in that the
  main diagonal is multiplied by $\alpha^0 = 1$, the first upper diagonal is
  multiplied by $\alpha^1 = 2$, the second upper diagonal is multiplied by
  $\alpha^2 = 1$, the first lower diagonal is multiplied by $\alpha^{-1} = 2$,
  and the second lower diagonal is multiplied by $\alpha^{-2} = 1$.
\end{Example}

We collect some remarks about the UDMs constructed in
Prop.~\ref{prop:udms:construction:1}.
\begin{itemize}

  \item All matrices $\matr{A}_{\ell}$, $2 \leq \ell < L$, are upper
    triangular matrices with non-zero diagonal entries. This follows from the
    fact that ${t \choose i} = 1$ if $t = i$ and ${t \choose i} = 0$ if $t <
    i$.

  \item The matrix $\matr{A}_2$ is an upper triangular matrix where the
    non-zero part equals Pascal's triangle (modulo $p$), see e.g.~$\matr{A}_2$
    in Ex.~\ref{ex:udm:2}. However, whereas usually Pascal's triangle is
    depicted such that the lines correspond to the upper entry in the binomial
    coefficient, here the vertical lines of the matrix correspond to the upper
    entry in the binomial coefficient.

  \item For $t \in [n]$, let us define the matrix $\mDelta_{t}$ of size $n
    \times n$: all entries are zero except $[\mDelta_{t}]_{t',t'} = +1$ for
    all $t' \in [n]$ and $[\mDelta_{t}]_{t'-1,t'} = -1$ for all $t < t' \leq
    n-1$. (Note that $\mDelta_{n-1} = \matr{I}_n$.) Because $\mDelta_{t}$ is
    an upper triangular matrix with non-zero diagonal entries it is an
    invertible matrix. One can show that $\matr{A}_2 \cdot \mDelta_0 \cdot
    \cdots \cdot \mDelta_{n-1} = \matr{I}_n$. Therefore, $\matr{A}_2 =
    \mDelta_{n-1}^{-1} \cdot \cdots \cdot \mDelta_0^{-1}$. Without going into
    the details, these $\mDelta_{t}$ matrices can be used (as part of the
    matrices needed) to solve the equation system $\matr{A} \cdot \vect{\hat
    u} = \vect{y}$ in Sec.~\ref{sec:introduction:1} with a type of Gaussian
    elimination.

  \item Applying Lemma~\ref{lemma:reduction:1} to $(q+1,n,q)$-UDMs as
    constructed in Prop.~\ref{prop:udms:construction:1} yields
    $(q+1,n-1,q)$-UDMs as constructed in Prop.~\ref{prop:udms:construction:1}.

  \item The setup in Sec.~\ref{sec:introduction:1} can be generalized as
    follows. Instead of sending vectors $\vect{x}_{\ell}$ of length $n$ we can
    also send vectors of length $n'$ where $n'$ is any positive
    integer. Obviously, the matrices $\matr{A}_{\ell}$ are then of size $n'
    \times n$. Essentially all results in this paper also hold for this setup,
    except for statements that involve the invertibility of the
    $\matr{A}_{\ell}$ matrices. Moreover, the sets $\set{K}^{= n}_{L}$ and
    $\set{K}^{\geq n}_{L}$ have to be modified to account for the fact that $0
    \leq k_{\ell} \leq n'$.
 
    Let us briefly focus on the case $n' = 1$, which results in the problem
    mentioned in Sec.~\ref{sec:introduction:1} whose solution used MDS codes,
    in particular Reed-Solomon codes. We let $\vect{x}$ be the the stacked
    version of all $\vect{x}_{\ell}$ vectors. Because $\vect{x}_{\ell}$ has
    length one, the vector $\vect{x}$ has length $L$. Similarly, we define the
    length-$L$ vector $\vect{y}$. It is not difficult to see that for the
    construction in Prop.~\ref{prop:udms:construction:1} the vector $\vect{x}$
    is an element of a doubly-extended Reed-Solomon
    code~\cite{MacWilliams:Sloane:77, Blahut:02:1} of length $L$, dimension
    $n$, and minimum distance $\dmins = L-n+1$. Note that $k_{\ell}$ can only
    be zero or one and that the sum $\sum_{\ell \in [L]} k_{\ell}$ equals the
    number of non-erased symbols in $\vect{y}$. In this case the proof of the
    construction in Prop.~\ref{prop:udms:construction:1} is very simple since
    we do not have to worry if a root has multiplicity one or higher. Indeed,
    let us show that if all non-erased entries of $\vect{y}$ are equal to zero
    then we must have $u(\iL) = 0$. If $k_1 = 1$ then $\deg(u(\iL)) \leq
    n-2$. However, the other non-erased entries of $\vect{y}$ require that
    $u(\iL)$ has at least $\sum_{\ell \in [L] \setminus \{ 1 \}} k_{\ell} = n
    - k_1 = n - 1$ roots. This is a contradiction. If $k_1 = 1$ then
    $\deg(u(\iL)) \leq n-1$. However, the other non-erased entries of
    $\vect{y}$ require that $u(\iL)$ has at least $\sum_{\ell \in [L]
    \setminus \{ 1 \}} k_{\ell} = n - k_1 = n$ roots. Again, this is a
    contradiction and so $u(\iL) = 0$ as desired. This argument is essentially
    equivalent to the proof used for showing that $\dmins \geq L-n+1$ for the
    above-mentioned doubly-extended Reed-Solomon code. (Together with the
    Singleton bound $\dmins \leq L-n+1$ we get $\dmins = L-n+1$.)

  \item Besides the generalization mentioned in the previous paragraph, the
    setup in Sec.~\ref{sec:introduction:1} can also be be generalized in the
    following way. Instead of requiring that decoding is uniquely possible for
    any $(k_0, \ldots, k_{L-1}) \in \set{K}^{\geq n}_L$ one may ask that
    decoding is uniquely possible for any $(k_0, \ldots, k_{L-1}) \in
    \set{K}^{\geq n''}_L$ where $n'' \geq n$. Of course, UDMs designed for $n''
    = n$ can be used for any $n'' \geq n$, however, for suitably chosen UDMs
    the required field size might be smaller, i.e.~$L \leq q+1$
    (cf.~Lemma~\ref{lemma:maximal:L:1}) might not be a necessary condition
    anymore. Indeed, in the same way as Goppa codes / algebraic-geometry
    codes~\cite{Stichentoth:93:1} are generalizations of Reed-Solomon codes,
    one can construct UDMs that are generalizations of the UDMs in
    Prop.~\ref{prop:udms:construction:1}. The generalization goes as follows:
    instead of obtaining the entries of the $\vect{x}_{\ell}$, $\ell \in [L]$,
    by evaluating the information polynomial (see
    Sec.~\ref{sec:proof:prop:udms:construction:1} for notation) at the
    rational points of the curve $L^q - L = 0$ (projectively: $L^q \tilde L -
    L \tilde L^q = 0$), they are obtained by evaluating the information
    polynomial at the rational places of a projective, geometrically
    irreducible, non-singular algebraic curve of genus $g \defeq n'' - n$. The
    proof for this setup is very similar to the proof in
    Sec.~\ref{sec:proof:prop:udms:construction:1}, however instead of the
    fundamental theorem of algebra one needs the Riemann-Roch
    theorem~\cite{Stichentoth:93:1}. Using Hasse-Weil-Serre
    bound~\cite{Stichentoth:93:1} one can generalize the result in
    Lemma~\ref{lemma:maximal:L:1} to the necessary condition $L \leq q + 1 +
    \lfloor 2 \sqrt{q} \rfloor g$. (Obviously, better bounds than the
    Hasse-Weil-Serre bound, cf.~e.g.~\cite{vanderGeer:vanderVlugt:05:1}, lead
    to better necessary conditions on $L$.)

  \item There is some connection between the construction in
    Prop.~\ref{prop:udms:construction:1} and so-called repeated-root cyclic
    codes~\cite{Chen:69:1, Castagnoli:Massey:Schoeller:vonSeemann:91:1,
    Lint:91:1, MorelosZaragoza:91:1}. Namely, the ``= 0'' part of
    Lemma~\ref{lemma:evaluation:1} is used to construct parity-check equations
    (and therefore a parity-check matrix) for a repeated-root cyclic code
    whose generator polynomial is
    known~\cite{Castagnoli:Massey:Schoeller:vonSeemann:91:1}.

\end{itemize}

\begin{Corollary}
  \label{cor:udms:construction:2}

  Consider the setup of Prop.~\ref{prop:udms:construction:1}. Let $p$ be the
  characteristic of $\GF{q}$, let $m$ be the smallest integer such that $n
  \leq p^m$, and let
  \begin{align*}
    i
      &= i_{m-1} p^{m-1} + \cdots + i_1 p + i_0,
    \quad
    0 \leq i_{h} < p, \
    h \in [m]
    \quad \text{and} \\
    t
      &= t_{m-1} p^{m-1} + \cdots + t_1 p + t_0,
    \quad
    0 \leq t_{h} < p, \ 
    h \in [m]
  \end{align*}
  be the radix-$p$ representations of $i \in [n]$ and $t \in [n]$,
  respectively. Then the entries of $\matr{A}_{\ell+2}$, $\ell \in [L-2]$, can
  be written as
  \begin{align*}
    [\matr{A}_{\ell+2}]_{i,t}
      &= \prod_{h \in [m]}
           {t_{h} \choose i_{h}}
           \alpha^{\ell (t_{h} - i_{h}) p^{h}}.
  \end{align*}
  This shows that in the case $n = p^m$ the matrices $\matr{A}_{\ell}$, $\ell
  \in [L]$ can be written as tensor products of some $p \times p$ matrices. In
  the special case $q = p$ (i.e.~$q$ is a prime) we can say more. Namely,
  letting $\matr{A}'_0, \ldots, \matr{A}'_{L-1}$ be the $(p+1,p,p)$-UDMs as
  constructed in Prop.~\ref{prop:udms:construction:1} we see that
  $\matr{A}_{\ell} = (\matr{A}'_{\ell})^{\otimes m}$ for all $\ell \in [L]$.
\end{Corollary}

\begin{Proof}
  Note that ${t \choose k}$ is an integer and therefore (by the natural
  mapping) an element of the prime subfield $\GF{p}$ of $\GF{q}$. Using the
  Lucas correspondence theorem which states that ${t \choose i} = \prod_{h \in
  [m]} {t_h \choose i_h}$ in $\GF{p}$ (and therefore also in $\GF{q}$), we
  obtain the reformulation. The last statement in the corollary follows from
  the fact that $\alpha^p = \alpha$ if $q = p$. (Note that for $\matr{A}_{0} =
  \matr{I}_{p^m}$ and $\matr{A}_{1} = \matr{J}_{p^m}$ it is trivial to verify
  that they can be written as tensor product and tensor powers of $p \times p$
  matrices.)
\end{Proof}

\mbox{}

Consider the same setup as in Cor.~\ref{cor:udms:construction:2}. Because $0
\leq i_h < p$, we observe that ${t_{h} \choose i_{h}}$ is a polynomial
function of degree $i_{h}$ in $t$. Using
Lemma~\ref{lemma:replace:and:add:lines:1}, the matrices can therefore be
modified so that the entries are
\begin{align*}
  [\matr{A}_{\ell+2}]_{i,t}
    &= \prod_{h \in [m]}
         t_{h}^{i_{h}} \alpha^{\ell (t_{h} - i_{h}) p^{h}},
         \quad
         (\ell,i,t) \in [L-2] \times [n] \times [n].
\end{align*}
Letting $q = p \defeq 2$, $n = 2^m$, $L \defeq q + 1 = 3$, and $\alpha \defeq
1$ we have $[\matr{A}_2]_{i,t} = \prod_{h \in [m]} t_{h}^{i_{h}}$, which
recovers the $(L{=}3,n{=}2^m,q{=}2)$-UDMs
in~\cite[Sec.~4.5.3]{Tavildar:Viswanath:05:1} since the latter matrix is a
Hadamard matrix. In general (i.e.~not just in the case $q = 2$), the fact that
the entries of $[\matr{A}_2]_{i,t}$ can be written as $[\matr{A}_2]_{i,t} =
\prod_{h \in [m]} t_{h}^{i_{h}}$, reminds very strongly of Reed-Muller
code~\cite{MacWilliams:Sloane:77, Blahut:02:1}. In the former case, the rows
of $\matr{A}_2$ are the evaluation of the multinomial function $(t_0, \ldots,
t_{m-1}) \mapsto \prod_{h \in [m]} t_{h}^{i_{h}}$, in the latter case the rows
of the generator matrix can be seen as the evaluation of multinomials at
various places.

Recall the $(L{=}4,n{=}3,q{=}3)$-UDMs $\matr{A}_0, \ldots, \matr{A}_3$ from
Ex.~\ref{ex:udm:2}. The authors of~\cite{Tavildar:Viswanath:05:1, Doshi:05:1}
conjecture that the tensor powers $\matr{A}_0^{\otimes m}, \ldots,
\matr{A}_3^{\otimes m}$ are $(4,3^m,3)$-UDMs for any positive integer
$m$. This is indeed the case and can be shown as follows. From
Ex.~\ref{ex:udms:construction:1} we know that $\matr{A}_0, \ldots,
\matr{A}_3$ can be obtained by the construction in
Prop.~\ref{prop:udms:construction:1}. Because $q = 3$ is a prime,
Cor.~\ref{cor:udms:construction:2} yields the desired conclusion that the
tensor powers $\matr{A}_0^{\otimes m}, \ldots, \matr{A}_3^{\otimes m}$ are
$(4,3^m,3)$-UDMs for any positive integer $m$.

\section{Conclusions}
\label{sec:conclusions:1}

We have presented an explicit construction of UDMs for all parameters $L$, $n$,
$q$ for which UDMs can potentially exist. They are essentially based on
Pascal's triangle (and modifications thereof) and the proof was heavily based
on properties of Hasse derivatives. We have also pointed out connections to
Reed-Solomon codes, Reed-Muller codes, and repeated-root cyclic codes. One
wonders if there are also other UDMs constructions that are not simply
reformulations of the present UDMs.

\appendix

\section{Proofs}

\label{sec:proofs:1}

\subsection{Proof of Lemma~\ref{lemma:reversed:order:1}}

\label{sec:proof:lemma:reversed:order:1}

It is sufficient to show how $\matr{A}_0$ and $\matr{A}_1$ can be used to
construct matrices $\matr{A}'_0$ and $\matr{A}'_1$ such that $\matr{A}'_0,
\matr{A}'_1, \matr{A}_2, \ldots, \matr{A}_{L-1}$ are $(L,n,q)$-UDMs and such
that $\matr{A}'_1$ is the same as $\matr{A}'_0$ except that the rows are in
reversed order. We use the following algorithm:
\begin{itemize}
  
  \item Assign $\matr{A}'_0 := \matr{A}_0$ and $\matr{A}'_1 := \matr{A}_1$.

  \item For $i$ from $0$ to $n-1$ do
    \begin{itemize}
    
      \item Let $\matr{B}'_0$ be the $(i+1) \times n$ matrix that contains the
        rows $0$ to $i$ from $\matr{A}'_0$. Similarly, let $\matr{B}'_1$ be
        the $(n-i) \times n$ matrix that contains the rows $0$ to $n-i-1$ from
        $\matr{A}'_0$.

      \item Build the $(n+1) \times n$-matrix $\matr{B}$ by stacking
        $\matr{B}'_0$ and $\matr{B}'_1$.

      \item Because of the size of $\matr{B}$, the left null space of
        $\matr{B}$ is non-empty. (In fact, because of the UDMs conditions the
        matrix $\matr{B}$ must have rank $n$ which implies that the left null
        space is one-dimensional.) Pick a non-zero (row) vector $\vect{b}^\tr$
        in this left null space, i.e.~$\vect{b}^\tr$ fulfills $\vect{b}^\tr
        \cdot \matr{B} = \vect{0}^\tr$. Write $\vect{b}^\tr = (
        {\vect{b}'_0}^\tr \, | \, {\vect{b}'_1}^\tr)$ where $\vect{b}'_0$ is
        of length $i+1$ and $\vect{b}'_1$ is of length $n-i$.

      \item Because of the UDMs conditions it can be seen that neither
        $[\vect{b}]_i$ nor $[\vect{b}]_{n+1}$ can be zero, i.e.~neither the
        last component of $\vect{b}'_0$ nor the last component of
        $\vect{b}'_1$ is zero. Replace the $i$-th row of matrix $\matr{A}'_0$
        by the vector ${\vect{b}'_0}^\tr \matr{B}'_0$. Similarly, replace the
        $(n-i-1)$-th row of matrix $\matr{A}'_1$ by the vector
        $-{\vect{b}'_1}^\tr \matr{B}'_1$. We see that the $i$-th row of
        $\matr{A}'_0$ equals the $(n-i-1)$-th row of $\matr{A}'_1$ and because
        of Lemma~\ref{lemma:replace:and:add:lines:1} the matrices
        $\matr{A}'_0, \matr{A}'_1, \matr{A}_2, \ldots, \matr{A}_{L-1}$ are
        still $(L,n,q)$-UDMs

    \end{itemize}

\end{itemize}
Applying the algorithm to $\matr{A}_2$ and $\matr{A}_3$, then to $\matr{A}_4$
and $\matr{A}_5$, $\ldots$ yields the desired result.

\subsection{Proof of Lemma~\ref{lemma:reduction:1}}

\label{sec:proof:lemma:reduction:1}

It is clear that $\matr{A}'_0 = \matr{I}_{n-1}$ and $\matr{A}'_1 =
\matr{J}_{n-1}$. It is enough to focus on the case $L > 2$ since for $L
\leq 2$ the lemma statement is easily verified.

So, fix some $L > 2$. We know that for any $(k_0, \ldots, k_{L-1}) \in
\set{K}^{= n}_L$ the UDMs condition is fulfilled for the matrices
$\matr{A}_0, \ldots, \matr{A}_{L-1}$. We have to show that for any $(k'_0,
\ldots, k'_{L-1}) \in \set{K}^{= n-1}_L$ the UDMs condition is also
fulfilled for the matrices $\matr{A}'_0, \ldots, \matr{A}'_{L-1}$.

Take such an $L$-tuple $(k'_0, \ldots, k'_{L-1}) \in \set{K}^{= n-1}_L$. If
$k'_{\ell} = 0$ for $2 \leq \ell < L$ then $k'_0 + k'_1 = n$ and it is clear
that the UDMs condition is fulfilled. So, assume that there is at least one
$\ell$ with $2 \leq \ell \leq L$ such that $k_{\ell} > 0$ (which implies among
other things that $k'_0 + k'_1 < n$). The $(n-1) \times (n-1)$-matrix
$\matr{A}'$ for which we have to check the full-rank condition looks like
\begin{align*}
  \matr{A}'
    &= \begin{pmatrix}
         \matr{I}_{k'_0} & \matr{0}   & \matr{0} \\
         \matr{0}        & \matr{0}   & \matr{J}_{k'_1} \\
         \matr{B}'       & \matr{B}'' & \matr{B}'''
       \end{pmatrix},
\end{align*}
where $\matr{B}'$, $\matr{B}''$, and $\matr{B}'''$ are matrices of size
$(n{-}k'_0{-}k'_1) \times k'_0$, $(n{-}k'_0{-}k'_1) \times (n{-}k'_0{-}k'_1)$,
and $(n{-}k'_0{-}k'_1) \times k'_1$, respectively, and where $[\matr{B}',
\matr{B}'', \matr{B}''']$ consists of rows from $\matr{A}'_{\ell}$, $2 \leq
\ell < L$. It can easily be seen that the $(n-1) \times (n-1)$-matrix 
$\matr{A}'$ has full rank if and only if the $n \times n$-matrix
\begin{align*}
  \matr{A}
    &= \begin{pmatrix}
         \matr{I}_{k'_0} & \matr{0}   & \matr{0}        & \vect{0} \\
         \matr{0}        & \matr{0}   & \matr{0}        & 1 \\
         \matr{0}        & \matr{0}   & \matr{J}_{k'_1} & \vect{0} \\
         \matr{B}'       & \matr{B}'' & \matr{B}'''     & \vect{b}
       \end{pmatrix}
     = \begin{pmatrix}
         \matr{I}_{k'_0} & \matr{0}   & \matr{0} \\
         \matr{0}        & \matr{0}   & \matr{J}_{k'_1+1} \\
         \matr{B}'       & \matr{B}'' & \matr{B}''''
       \end{pmatrix}
\end{align*}
has full rank, where $\vect{b}$ is an arbitrary length-$(n{-}k'_0{-}k'_1)$
vector and where $\matr{B}'''' \defeq [\matr{B}''' \ | \ \vect{b}]$.

Let $k_{\ell} \defeq k'_{\ell}$ for $\ell \in [L] \setminus \{ 1 \}$ and let
$k_1 \defeq k'_1 + 1$. (Because $\sum_{\ell \in [L]} k'_{\ell} = n-1$ we have
$\sum_{\ell \in [L]} k_{\ell} = n$.) Choosing $\vect{b}$ such that the first
$k_2$ entries of $\vect{b}$ equal the top $k_2$ entries of the $(n-1)$-th
column of $\matr{A}_2$, $\ldots$, the last $k_{L-1}$ entries of $\vect{b}$
equal the top $k_{L-1}$ entries of the $(n-1)$-th column of $\matr{A}_{L-1}$,
we see that $\matr{A}$ represents the matrix that we have to look at when
checking the UDMs property for $(k_0, \ldots, k_{L-1})$ for $\matr{A}_0,
\ldots, \matr{A}_{L-1}$. However, by assumption we know that $\matr{A}$ has 
full rank and so the matrix $\matr{A}'$ has also full rank.

\subsection{Proof of Lemma~\ref{lemma:maximal:L:1}}

\label{sec:proof:lemma:maximal:L:1}

Assume that $\matr{A}_0, \ldots, \matr{A}_{L-1}$ are $(L,n,q)$-UDMs. The
comments after Lemma~\ref{lemma:reversed:order:1} allows to assume without
loss of generality that $\matr{A}_0 = \matr{I}_n$ and that $\matr{A}_1 =
\matr{J}_n$.

First, we want to show that all entries in the first row of $\matr{A}_{\ell}$,
$2 \leq \ell < L$, must be non-zero. Indeed, for $2 \leq \ell < L$ and $m \in
[n]$ the UDMs condition for $k_0 = m$, $k_1 = n - m - 1$, and $k_{\ell} = 1$
(all other $k_{\ell'}$ are zero) shows that $[\matr{A}_{\ell}]_{0,m} \neq
0$. Using Lemma~\ref{lemma:replace:and:add:lines:1} we can therefore without
loss of generality assume that $[\matr{A}_{\ell}]_{0,n-1} = 1$ for all $2 \leq
\ell < L$.

Secondly, the UDMs condition for $k_0 = n-2$, $k_{\ell} = 1$, and $k_{\ell'} =
1$ (all other $k_{\ell''}$ are zero) implies that the matrix
\begin{align*}
  \begin{pmatrix}
    [\matr{A}_{\ell}]_{0,n-2}  & [\matr{A}_{\ell}]_{0,n-1} \\
    [\matr{A}_{\ell'}]_{0,n-2} & [\matr{A}_{\ell'}]_{0,n-1}
  \end{pmatrix}
    &= \begin{pmatrix}
         [\matr{A}_{\ell}]_{0,n-2}  & 1 \\
	 [\matr{A}_{\ell'}]_{0,n-2} & 1
       \end{pmatrix}
\end{align*}
must have rank $2$ for any distinct $\ell$ and $\ell'$ fulfilling $2 \leq \ell
< L$ and $2 \leq \ell' < L$. It is not difficult to see that this implies that
$[\matr{A}_{\ell}]_{0,n-2}$ must be distinct for all $2 \leq \ell < L$. Since
$[\matr{A}_{\ell}]_{0,n-2}$ must be non-zero and since $\GF{q}$ has $q-1$
non-zero elements we see that $L-2 \leq q-1$, i.e.~$L \leq q+1$.

\subsection{Proof of Proposition~\ref{prop:udms:construction:1}}

\label{sec:proof:prop:udms:construction:1}

We will use the following notation. We set $\beta_0 \defeq 0$, $\beta_{\ell+2}
\defeq \alpha^{\ell}$, $\ell \in [L-2]$. Because $\alpha$ is a primitive 
element of $\GF{q}$, all $\beta_i$'s are distinct. (Note that $\beta_1$ has
not been defined.) Let $u_t \defeq [\vect{u}]_t$, $t \in [n]$, where
$\vect{u}$ is the information vector, cf.~Sec.~\ref{sec:introduction:1}, and
let the information polynomial be the polynomial
\begin{align*}
  u(\iL)
    &\defeq
       \sum_{t \in [n]}
         u_t \iL^t \in \GF{q}[\iL],
\end{align*}
whose degree $\deg(u(\iL))$ is at most $n-1$. Moreover, we let $\matr{X}
\defeq [\vect{x}_0 | \cdots | \vect{x}_{L-1}]$ be a matrix of size 
$n \times L$ with entries $x_{i,\ell} \defeq [\matr{X}]_{i,\ell} =
[\vect{x}_{\ell}]_i$ for $(i,\ell) \in [n] \times [L]$. The $n \times
L$-matrix $\matr{Y}$ is defined similarly.

\begin{Lemma}
  \label{lemma:a:matrix:entries:1}

  Using Hasse derivatives, the elements of $\matr{A}_{\ell}$, $\ell \in [L]
  \setminus \{ 1 \}$, can be expressed as
  \begin{align*}
    [\matr{A}_{\ell}]_{i,t}
      &= \left.
           \HD{\iL}{i}
             \big(
               \iL^t
             \big)
         \right|_{\iL = \beta_{\ell}}
         \quad
         (\ell \in [L] \setminus \{ 1 \}, \ i \in [n], \ t \in n).
  \end{align*}
\end{Lemma}

\begin{Proof}
  Applying the definition of the Hasse derivative
  (cf.~Sec.~\ref{sec:hasse:derivative:1}) we get $\HD{\iL}{i} \big( \iL^t
  \big) = {t \choose i}\iL^{t-i}$, where ${t \choose i} \iL^{t-i}$ is the zero
  polynomial if $t < i$. Upon substituting $\iL = \beta_{\ell}$ we obtain
  \begin{align*}
    \left.
      \HD{\iL}{i}
        \big(
          \iL^t
        \big)
    \right|_{\iL = \beta_{\ell}}
      &= {t \choose i}
         \beta_{\ell}^{t-i}
       = \left\{
           \begin{array}{ll}
             1                                
               & (\ell = 0, \ t = i)  \\
             0                                
               & (\ell = 0, \ t \neq i) \\
             {t \choose i} \alpha^{(\ell-2)(t-i)} 
               & (\ell \in \{ 2, \ldots, L-1 \}
           \end{array}
         \right\}
       = [\matr{A}_{\ell}]_{i,t}.
  \end{align*}
  
\end{Proof}

\begin{Lemma}
  \label{lemma:y:vectors:entries:1}

  The elements of $\matr{X}$ can be expressed as
  \begin{align*}
    x_{i,\ell}
      &= \begin{cases}
           \left.
             \HD{\iL}{i}
               \big(
                 u(\iL)
               \big)
           \right|_{\iL = \beta_{\ell}}
             & (i \in [n], \ \ell \in [L] \setminus \{ 1 \}) \\
           u_{n-1-i}
             & (i \in [n], \ \ell = 1)
         \end{cases}.
  \end{align*}
\end{Lemma}

\begin{Proof}
  Remember that in Sec.~\ref{sec:introduction:1} we defined $\vect{x}_{\ell} =
  \matr{A}_{\ell} \cdot \vect{u}$. The result for $\ell = 1$ is clear. For
  $\ell \in [L] \setminus \{ 1 \}$ we use the results of
  Lemma~\ref{lemma:a:matrix:entries:1} and the linearity of the Hasse
  derivative to obtain
  \begin{align*}
    x_{i, \ell}
      &= [\vect{x}_{\ell}]_i
       = \sum_{t \in [n]}
           [\matr{A}_{\ell}]_{i,t}
           [\vect{u}]_t
       = \left.
           \sum_{t \in [n]}
             u_t
             \HD{\iL}{i}
               \big(
                 \iL^t
               \big)
         \right|_{\iL = \beta_{\ell}}
       = \left.
           \HD{\iL}{i}
             \left(
               \sum_{t \in [n]}
                 u_t
		   \iL^t
             \right)
         \right|_{\iL = \beta_{\ell}}
  \end{align*}
  for all $(i,\ell) \in [n] \times ([L] \setminus \{ 1 \})$.
\end{Proof}

After these preliminary lemmas, let us turn to task of checking the UDMs
condition for all $(k_0, \ldots, k_{L-1}) \in \set{K}^{= n}_L$. Fix such a
tuple $(k_0, \ldots, k_{L-1}) \in \set{K}^{= n}_L$ and let $\psi$ be the
mapping of the vector $\vect{u}$ to the non-erased entries of the matrix
$\matr{Y}$; it is clear that $\psi$ is a linear mapping. Reconstructing
$\vect{u}$ is therefore nothing else than applying the mapping $\psi^{-1}$ to
the non-erased positions of $\matr{Y}$. However, this gives a unique vector
$\vect{u}$ only if $\psi$ is an injective function. Because $\psi$ is linear,
showing injectivity of $\psi$ is equivalent to showing that the kernel of
$\psi$ contains only the vector $\vect{u} = \vect{0}$, or equivalently, only
the polynomial $u(\iL) = 0$.

So, let us show that the only possible pre-image of 
\begin{align*}
  y_{i,\ell}
    &= 0,
       \quad
       i \in [k_{\ell}], \ \ell \in [L],
\end{align*}
or, equivalently, of
\begin{align*}
  x_{i,\ell}
    &= 0,
       \quad
       i \in [k_{\ell}], \ \ell \in [L],
\end{align*}
is $u(\iL) = 0$. Using Lemma~\ref{lemma:y:vectors:entries:1}, this is
equivalent to showing that
\begin{alignat}{2}
  \left.
    \HD{\iL}{i}
      \big(
        u(\iL)
      \big)
  \right|_{\iL = \beta_{\ell}}
    &= 0
       \quad 
       &&(i \in [k_{\ell}], \ \ell \in [L] \setminus \{ 1 \})
         \label{eq:zero:hasse:derivatives:1} \\
  u_{n-1-i}
    &= 0
       \quad
       && (i \in [k_1])
         \label{eq:u:zero:positions:1}
\end{alignat}
implies that $u(\iL) = 0$. In a first
step,~\eqref{eq:zero:hasse:derivatives:1} together with
Cor.~\ref{cor:evaluation:2} tell us that $\beta_{\ell}$, $\ell \in [L]
\setminus \{ 1 \}$, must be a root of $u(\iL)$ of multiplicity at least
$k_{\ell}$. Adding up and using the fundamental theorem of algebra we get
\begin{align}
  \deg(u(\iL))
    &\geq
       \sum_{\ell \in [L] \setminus \{ 1 \}}
         k_\ell
     = n - k_1
  \quad\quad
  \text{ or }
  \quad\quad
  u(\iL) = 0.
         \label{eq:lower:bound:number:of:roots:1}
\end{align}
In a second step, ~\eqref{eq:u:zero:positions:1} tells us that we must have
$\deg(u(\iL)) \leq n - 1 - k_1$. Combining this
with~\eqref{eq:lower:bound:number:of:roots:1}, we obtain the desired result
that $u(\iL) = 0$.

In our proof, the matrix $\matr{A}_1$ and the vector $\vect{x}_1$ had a
special position. On wonders if it is possible to homogenize the setup so as
to compactify the notation. Something like this is indeed possible. Letting
$\mathbb{P}_{\GF{q}}^1 \defeq \GF{q} \cup \{ \infty \}$ be the projective line
over $\GF{q}$, the matrix $\matr{A}_1$ and the vector $\vect{y}_1$ correspond
so-to-speak to the point $\infty$. More precisely, let
\begin{align*}
  u(\iL,\tilde \iL)
    &\defeq
       \sum_{t \in [n]}
         u_t \iL^t \tilde \iL^{n-1-t} \in \GF{q}[\iL],
\end{align*}
be the homogenized information polynomial. Whereas setting $(\iL, \tilde \iL)
\defeq (\iL,1)$ gives the original information polynomial, setting 
$(\iL, \tilde \iL) \defeq (1,0)$ corresponds to evaluating the original
information polynomial at $\infty$. In formulas,
Lemmas~\ref{lemma:a:matrix:entries:1} and~\ref{lemma:y:vectors:entries:1} read
now
\begin{align*}
  [\matr{A}_{\ell}]_{i,t}
    &= \begin{cases}
         \left.
           \HD{\iL}{i}
             \big(
               \iL^t \tilde \iL^{n-1-t}
             \big)
         \right|_{(\iL, \tilde \iL) = (\beta_{\ell},1)}
         & (\ell \in [L] \setminus \{ 1 \}, \ i \in [n], \ t \in n) \\
         \left.
           \HD{\tilde \iL}{i}
             \big(
               \iL^t \tilde \iL^{n-1-t}
             \big)
         \right|_{(\iL, \tilde \iL) = (1,0)}
         & (\ell = 1, \ i \in [n], \ t \in n) \\
       \end{cases}
\end{align*}
and
\begin{align*}
  x_{i,\ell}
    &= \begin{cases}
         \left.
           \HD{\iL}{i}
             \big(
               u(\iL, \tilde \iL)
             \big)
         \right|_{(\iL, \tilde \iL) = (\beta_{\ell},1)}
           & (i \in [n], \ \ell \in [L] \setminus \{ 1 \}) \\
         \left.
           \HD{\tilde \iL}{i}
             \big(
               u(\iL, \tilde \iL)
             \big)
         \right|_{(\iL, \tilde \iL) = (1,0)}
           & (i \in [n], \ \ell = 1)
       \end{cases}.
\end{align*}

\section{The Hasse Derivative}

\label{sec:hasse:derivative:1}

The Hasse derivative was introduced in~\cite{Hasse:36:1}. Throughout this
appendix, let $q$ be some prime power. For any non-negative integer $i$, the
$i$-th Hasse derivative of a polynomial $\sum_{k=0}^{d} a_k \iX^k \in
\GF{q}[\iX]$ is defined to be\footnote{The $i$-th formal derivative equals
$i!$ times the Hasse derivative: so, for fields with characteristic zero there
is not a big difference between these two derivatives since $i!$ is always
non-zero, however for finite fields there can be quite a gap between these two
derivatives since $i!$ can be zero or non-zero.}
\begin{align*}
  \HD{\iX}{i}
    \left(
      \sum_{k=0}^{d}
        a_k \iX^k
    \right)
      &\defeq
         \sum_{k=0}^{d}
           {k \choose i}  a_k \iX^{k-i}.
\end{align*}
Note that when $i > k$ then ${k \choose i} \iX^{k-i} = 0$, i.e.~the zero
polynomial. We list some well-know properties of the Hasse derivative:
\begin{align*}
  \HD{\iX}{i}
    \big(
      \gamma f(\iX) + \eta g(\iX)
    \big)
      &= \gamma \HD{\iX}{i}\big( f(\iX) \big)
         +
         \eta \HD{\iX}{i}\big( g(\iX) \big), \\
  \HD{\iX}{i}
    \big(
      f(\iX) g(\iX)
    \big)
      &= \sum_{i'=0}^{d}
           \HD{\iX}{i'}\big( f(\iX) \big)
           \HD{\iX}{i-i'}\big( g(\iX) \big), \\
  \HD{\iX}{i}
    \left(
      \prod_{h \in [M]}
        f_{h}(\iX)
    \right)
      &= \sum_{(i_0, \ldots, i_{M-1}) \in \set{K}^{= i}_M} \
           \prod_{h \in [M]}
             \HD{\iX}{i_{h}}\big( f_{h}(\iX) \big), \\
  \HD{\iX}{i}
    \left(
      (X-\gamma)^k
    \right)
      &= {k \choose i} (X-\gamma)^{k-i},
\end{align*}
where $k$ and $i$ are some non-negative integers, $M$ is some positive integer,
and where $\gamma, \eta \in \GF{q}$. Be careful that $\HD{\iX}{i_1}
\HD{\iX}{i_2} \neq \HD{\iX}{i_1 + i_2}$ in general. However, it holds that
$\HD{\iX}{i_1} \HD{\iX}{i_2} = {i_1 + i_2 \choose i_1} \HD{\iX}{i_1 + i_2}$.

\begin{Lemma}
  \label{lemma:evaluation:1}

  Let $q$ be some prime power and let us denote the elements of $\GF{q}$ by
  $\gamma_r$, $r \in [q]$, i.e.~$\GF{q} \defeq \{ \gamma_0, \ldots,
  \gamma_{q-1} \}$. If $m_0, \ldots, m_{q-1}$ are some non-negative integers
  then for any $r \in [q]$ we have
  \begin{align*}
    \left.
      \HD{\iX}{i}
        \left(
          \prod_{r' \in [q]}
            (X - \gamma_{r'})^{m_{r'}}
        \right)
    \right|_{\iX = \gamma_r}
        &\quad
         \begin{cases}
           = 0    & (0 \leq i < m_r) \\
           \neq 0 & (i = m_r) 
         \end{cases}
  \end{align*}
\end{Lemma}

\begin{Proof}
  Using properties of the Hasse derivative we see that
  \begin{align*}
    \left.
      \HD{\iX}{i}
        \left(
          \prod_{r' \in [q]}
            (X - \gamma_{r'})^{m_{r'}}
        \right)
    \right|_{\iX = \gamma_r}
        &= \left.
             \sum_{(i_0, \ldots, i_{q-1}) \in \set{K}^{= i}_q}
               \prod_{r' \in [q]}
                 \HD{\iX}{i_{r'}}
                   \big(
                     (X - \gamma_{r'})^{m_{r'}}
                   \big)
           \right|_{\iX = \gamma_r} \\
        &= \left.
             \sum_{(i_0, \ldots, i_{q-1}) \in \set{K}^{= i}_q}
               \prod_{r' \in [q]}
                 {m_{r'} \choose i_{r'}}
                 (X - \gamma_{r'})^{m_{r'} - i_{r'}}
           \right|_{\iX = \gamma_r} \\
        &= \sum_{(i_0, \ldots, i_{q-1}) \in \set{K}^{= i}_q}
             \prod_{r' \in [q]}
               {m_{r'} \choose i_{r'}}
               (\gamma_r - \gamma_{r'})^{m_{r'} - i_{r'}}
  \end{align*}
  The polynomial ${m_r \choose i_r} (X - \gamma_r)^{m_r - i_r}$ is the zero
  polynomial for $i_r > m_r$ and so ${m_r \choose i_r} (\gamma_r -
  \gamma_r)^{m_r - i_r}$ is interpreted to be zero if $i_r > m_r$. If $0 \leq
  i < m_r$ then $0 \leq i_r < m_r$ and so all summands are zero, yielding a
  zero sum. However, if $i = m_r$ there is exactly one summand that is
  non-zero, namely when $i_{r'} = 0$, $r' \in [q] \setminus \{ r \}$ and $i_r
  = m_r$ and so
  \begin{align*}
    \left.
      \HD{\iX}{m_r}
        \left(
          \prod_{r' \in [q]}
            (X - \gamma_{r'})^{m_{r'}}
        \right)
    \right|_{\iX = \gamma_r}
        &= {m_r \choose m_r}
           (\gamma_r - \gamma_r)^{m_r - m_r}
           \prod_{r' \in [q] \setminus \{ r \}}
             {k_{r'} \choose 0}
             (\gamma_r - \gamma_{r'})^{k_{r'} - 0} \\
        &= \prod_{r' \in [q] \setminus \{ r \}}
             (\gamma_r - \gamma_{r'})^{k_{r'}}
         \neq 0.
  \end{align*}

\end{Proof}

\begin{Corollary}
  \label{cor:evaluation:2}

  Let $p(\iX) \in \GF{q}[\iX]$. If for some $\beta \in \GF{q}$ and some
  non-negative integer $m$ it holds that
  \begin{align*}
    \left.
      \HD{\iX}{i}
        \big(
          p(\iX)
        \big)
    \right|_{\iX = \beta}
      &= 0
         \quad
         \text{for all $i \in [m]$},
  \end{align*}
  then $\beta$ is a root of $p(\iX)$ with multiplicity at least $m$.
\end{Corollary}

\begin{Proof}
  Let $q'$ be a power of $q$ such that the polynomial $p(\iX)$ splits in
  $\GF{q'} = \{ \gamma_0, \ldots, \gamma_{q'} \}$, i.e.~so that all roots of
  $p(\iX)$ are in $\GF{q'}$. (The theory of finite fields tells us that such a
  $q'$ always exists.) Then there are non-negative integers $m_0, \ldots,
  m_{q'}$ and a non-zero $\eta \in \GF{q}$ such that
  \begin{align}
    p(\iX)
      &= \eta
         \prod_{r' \in [q']}
           (\iX - \gamma_{r'})^{m_{r'}}
             \label{eq:splitting:of:p:1}
  \end{align}
  and such that $\sum_{r' \in [q']} m_{r'} = \deg(p(\iX))$. Let $r \in [q']$
  be such that $\beta = \gamma_r$. The proof will be by contradiction. So,
  assume for the moment that $m_r < m$, i.e.~that $\beta$ is a root of
  $p(\iX)$ of multiplicity $m_r$ smaller than $m$. Using
  Lemma~\ref{lemma:evaluation:1} and~Eq.~\eqref{eq:splitting:of:p:1} we see
  that
  \begin{align*}
    \left.
      \HD{\iL}{m_r}
        \big(
          p(\iX)
        \big)
    \right|_{\iX = \beta}
       \neq 0,
  \end{align*}
  which is a contradiction to the assumption made in the corollary
  statement. This proves the corollary.
\end{Proof}

{
\bibliographystyle{ieeetr}
\bibliography{/home/vontobel/references/references}
}

\end{document}